\documentclass[12pt]{iopart}

\expandafter\let\csname equation*\endcsname\relax
  \expandafter\let\csname endequation*\endcsname\relax

\usepackage{amsmath}
\usepackage{amssymb}
\usepackage{graphicx}
\usepackage{color}

\newcommand{\beq}{\begin{equation}}
\newcommand{\eeq}{\end{equation}}
\newcommand{\ben}{\begin{eqnarray}}
\newcommand{\een}{\end{eqnarray}}
\newcommand{\pa}{\partial}

\newcommand{\lap}{\Delta}

\newcommand{\bx}{\mathbf{x}}
\newcommand{\calf}{\mathcal{F}}
\newcommand{\cir}{\circlearrowleft}
\newcommand{\paz}{\pa_z}
\newcommand{\pao}{\pa_0}
\newcommand{\pau}{\pa_1}
\newcommand{\tf}{\tilde{f}}
\newcommand{\tpo}{\tilde{P}_0}
\newcommand{\tpu}{\tilde{P}_1}
\newcommand{\tpd}{\tilde{P}_2}
\newcommand{\tfc}{\tilde{\calf}}
\newcommand{\tc}{\tilde{c}}
\newcommand{\thot}{\tilde{h}}
\newcommand{\tx}{\tilde{x}}
\newcommand{\tvl}{\tilde{V}}
\newcommand{\cala}{\mathcal{A}}
\newcommand{\pax}{\partial_x}
\newcommand{\pay}{\partial_y}

\begin{document}

\title{Hamiltonian closures for two-moment fluid models derived from drift-kinetic equations}

\author{E. Tassi$^{1,2}$}
\address{$^1$ Aix-Marseille Universit\'e, CNRS, CPT, UMR 7332, 13288 Marseille, France \\
$^2$ Universit\'e de Toulon, CNRS, CPT, UMR 7332, 83957 La Garde, France}

\baselineskip 24 pt

\begin{abstract}
We derive the conditions under which the fluid models obtained from the first two moments of Hamiltonian drift-kinetic systems of interest to plasma physics, preserve a Hamiltonian structure. The adopted procedure consists of determining closure relations that allow to truncate the Poisson bracket of the drift-kinetic system, expressed in terms of the moments, in such a way that the resulting operation is a Poisson bracket for functionals of the first two fluid moments. The analysis is carried out for a class of full drift-kinetic equations and also for drift-kinetic systems in which a splitting between an equilibrium distribution function and a perturbation is performed. In the former case we obtain that the only closure, not involving integral or differential operators, that leads to a Poisson bracket, corresponds to that of an ideal adiabatic gas made of molecules possessing one degree of freedom. In the latter case, Hamiltonian closures turn out to be those in which the second moment is a linear combination of the first two moments, which can be seen as a linearization of the Hamiltonian closure of the full drift-kinetic case. A number of weakly-3D Hamiltonian reduced fluid models of interest, for instance for tokamak plasmas, can be derived in this way and, viceversa given a fluid model with a Hamiltonian structure of a certain type, a parent Hamiltonian drift-kinetic model can then be identified. We make use of this correspondence to identify the drift-kinetic models from which Hamiltonian fluid models for magnetic reconnection and compressible plasma dynamics in the presence of a static but inhomogeneous magnetic field can be derived. The Casimir invariants of the Poisson brackets of the derived fluid models are also discussed. It is also shown that the Poisson structure for the fluid model derived from the full drift-kinetic system coincides with that of a reduced fluid model, when using the fluid velocity instead of the momentum as dynamical variable.

\end{abstract}
\maketitle

\section{Introduction}

Truncating moment hierarchies of kinetic theories is a customary procedure for obtaining fluid models that describe the dynamics of media such as fluids and plasmas. In the non-dissipative limit, typically the parent kinetic theory consists of the Vlasov equation, which possesses a Hamiltonian structure, characterized by a Lie-Poisson bracket \cite{Mor80,Mar82}, possibly coupled with other equations describing the dynamics of other fields, a typical example being the Vlasov-Maxwell system. The Vlasov equation describes the evolution of a distribution function $f({\bf x}, {\bf v})$, with ${\bf x}$ and ${\bf v}$ denoting the space and velocity coordinate, respectively. As is well known, the Vlasov equation can be replaced by an infinite hierarchy of evolution equation for the kinetic moments $P_n^{ijk} ({\bf x})=\int d^3 v v_x^i v_y^j v_z^k f({\bf x},{\bf v})$, with $n, i, j$ and $k$ non-negative integers such that $i+j+k=n$. From the point of view of the dynamical properties, it was shown \cite{Gib81} that the resulting hierarchy of moment equations inherits a Hamiltonian structure \cite{Kup78} from the Lie-Poisson structure of the Vlasov equation. The case of the infinite BBGKY (Bogoliubov-Born-Green-Kirkwood-Yvon) has also been treated in this context, and its Lie-Poisson structure was presented in Ref. \cite{Mar84}. More recently,  truncated moment hierarchies of the geodesic Vlasov equation have been shown to be related to integrable systems\cite{Gib08} and a geometric interpretation of the Lie-Poisson structure associated with the dynamics of the moments of the Vlasov equation has been presented \cite{Gib08b}. 

In the physics of strongly magnetized plasmas, such as those of tokamak devices, a widely adopted kinetic theory, is that based on the so-called drift-kinetic equation (see, e.g. Refs. \cite{Haz03,Nis00}). Such equation governs the evolution of the distribution function $\mathsf{f}(x,y,z,v,\mu)$ of the guiding centers of the particles, whose individual dynamics takes place on a reduced phase space consisting of the three spatial coordinates $x$, $y$ and $z$, of the velocity coordinate $v$, which is is directed along the dominant component of the magnetic field, and of the magnetic moment $\mu$, which is in general an adiabatic invariant of the guiding center dynamics. The advantage of dealing with a distribution function on a reduced phase space is evident, for instance when numerical computations are required. This occurs at the expense of having averaged out the particle gyro motion, which, anyway, takes place on fast time scales of little interest for most applications.
Analogously to the ordinary Vlasov equation, the collisionless drift-kinetic equation also possesses a noncanonical Hamiltonian structure, which reflects the Hamiltonian dynamics of the individual guiding centers. An important property that reduced fluid models derived as a truncated moment hierarchy of drift-kinetic equations, should respect, is that, when no dissipative and/or forcing terms are voluntarily added, they also possess a Hamiltonian structure, like their parent model. 
If the hierarchy of fluid equations obtained from a Hamiltonian drift-kinetic (or, in general, kinetic) theory is such that the evolution equations for the first $n$ moments involves the first $n+k$ moments, then the customary procedure followed to derive a closed $n$-moment fluid model, is to express the moments of order $n+1,\cdots n+k$, in terms of the first $n$ moments. Imposing such closure relations, however, does not, in general, preserve the original Hamiltonian character of the underlying drift-kinetic theory.       
In this article we propose a way to derive, from a Hamiltonian drift-kinetic system, two-moment fluid models in such a way that a Hamiltonian structure is preserved. More precisely, we express the Poisson bracket of drift-kinetic systems in terms of the kinetic moments with respect to the parallel velocity, for a fixed value of $\mu$, and truncate the resulting expression by considering functionals of the first two moments. Then we look for closures such that, when inserted into the truncated bracket, yield a Poisson bracket, thus, guaranteeing the existence of a Hamiltonian structure for the fluid model. We refer to these as to Hamiltonian closures. We anticipate that, although the analysis is restricted to the first two moments, this is already sufficient to reveal qualitative differences with respect to the Vlasov case, for which such Hamiltonian closures have been investigated \cite{Mor06}. Indeed, as will be shown in Sec. \ref{sec:dk}, unlike the case of the Vlasov equation \cite{Gib08,DeG12}, for drift-kinetic systems the functionals of the first two moments do not form a sub-algebra with respect to the Poisson bracket of the drift-kinetic systems expressed in terms of the moments. 

We carry out the analysis for two classes of drift-kinetic systems: a first class, describing the evolution of the full distribution function, and a second class, in which, as is commonly adopted in plasma physics, one follows the evolution of the perturbation of an equilibrium distribution function which is spatially homogeneous. 

By means of such Hamiltonian closures we can then derive reduced fluid models which automatically possess a Hamiltonian structure, that can also be known explicitly. Indeed, we also identify the conditions that a generic Hamiltonian functional of the first two moments has to satisfy, in order for the dynamical equations generated by the truncated Poisson bracket, to match the fluid equations obtained directly by taking moments of the drift-kinetic system and imposing the Hamiltonian closure. This allows for an identification, by means of the Hamiltonian structure, of reduced fluid models which come from Hamiltonian drift-kinetic models and makes it possible, for such fluid models, to reconstruct the Hamiltonian parent drift-kinetic model, if so wished. 

The article is organized as follows. In Sec. \ref{sec:dk} the Hamiltonian closures for the fluid models obtained from the drift-kinetic systems involving the full distribution function are derived and discussed. Sec. \ref{sec:df} deals with the Hamiltonian closures for the perturbed drift-kinetic systems. In particular, Sec. \ref{ssec:corr}, discusses the correspondence between Hamiltonian weakly-3D fluid models and Hamiltonian perturbed drift-kinetic models. Two examples are treated in detail, and a third example concerning an extension to multiple species is also described. Sec. \ref{sec:concl} is devoted to conclusions, whereas \ref{appa} and \ref{appb} present the details of the calculations concerning the closures that allow to respect the Jacobi identity.

\section{Hamiltonian drift-kinetic model} \label{sec:dk}

We consider the following drift-kinetic equation
\beq  \label{dk}
\frac{\pa f}{\pa t}-q[\varphi,f]_x -[\mathcal{B} , f]_x+v \frac{\pa f}{\pa z}-\frac{q}{M}\frac{\pa \varphi}{\pa z}\frac{\pa f}{\pa v}=0,
\eeq
coupled with Poisson's equation 
\ben  \label{poiss}
\varphi (\bx)=L\int dv f(\bx , v).
\een
In (\ref{dk})-(\ref{poiss}), $f(\bx,v)$ is a guiding center distribution function depending on spatial variables $\bx \in \mathcal{D} \subset \mathbb{R}^3$ and on the velocity $v \in \mathbb{R}$ in the direction parallel to the dominant component of the magnetic field. The latter is taken to be $\mathbf{B}(x)=B(1-x/l) \hat{z}$, with constant $B$ and $l$, such that $x/l \ll 1$. This corresponds to the most simplistic choice that retains magnetic inhomogeneities in a strong magnetic field, but the subsequent analysis can be done in principle for more general magnetic field configurations. The  distribution function $f$ has to be interpreted as the result of an integration, over the magnetic moment coordinate, of the actual drift-kinetic distribution function $\mathsf{f} (\bx , v, \mu) $, which includes the dependence on the magnetic moment $\mu $. In particular, we assumed that $\mathsf{f}(\bx , v , \mu ') = \delta ( \mu ' - \mu) f(\bx , v)$, so that we are actually restricting to a particular value $\mu ' =\mu$.

The constants $q$ and $M$ indicate the charge and the particle mass of the species under consideration, whereas the function $\mathcal{B}(x)=\mu B(1-x/l)$ identifies the term associated with the gradient-$B$ drift.

The brackets $[ , ]_{x,v}$ are defined in the following way: 
\beq
[f,g]_x=-\frac{1}{qB}\left(\frac{\pa f}{\pa x}\frac{\pa g}{\pa y}-\frac{\pa f}{\pa y}\frac{\pa g}{\pa x}\right), \qquad  [f,g]_v=\frac{1}{M}\left(\frac{\pa f}{\pa z}\frac{\pa g}{\pa v}-\frac{\pa f}{\pa v}\frac{\pa g}{\pa z}\right).
\eeq
By virtue of this definition, one can then see that the second and third term on the left-hand side of Eq. (\ref{dk}) indicate the advection of the distribution function by the $\mathbf{E}\times \mathbf{B}$ and gradient-$B$ drifts, respectively. The fourth term represents the free streaming along the direction of the magnetic field, whereas the fifth term accounts the acceleration due to the electric field.
 
Poisson's equation (\ref{poiss}), on the other hand, relates the electrostatic potential $\varphi$ with the guiding center density, by means of the operator $L$.  At an abstract level, our theory applies in general if $L$ is a linear operator, in general of integral type, with coefficients independent on $v$, and symmetric with respect to the $L^2 (\mathcal{D})$ inner product. From a more physical point of view, although these hypotheses on the operator $L$ exclude the most general forms of drift-kinetic Poisson's equation \cite{Pfi84,Pfi85,Kau86,Bri07}, they allow for the use of some simplified, but still relevant forms. As a paradigmatic case, let us assume a distribution function for ion guiding centers with $q=e$, where $e$ is the unit charge, and consider as spatial domain $\mathcal{D}$ a unit cube, over which we assume that the potential $\varphi$ has zero mean value. Then we can take as Poisson's equation the following relation
\ben  \label{poiss1}
-\Delta_{\perp} \varphi +\frac{e^2 B^2}{M T_e} (\varphi - < \varphi>)=\frac{e B^2}{M n_0} \left(\int dv f - n_0 \right),
\een
where $\Delta_{\perp}$ indicates the Laplacian operator in the $xy$ plane, $T_e$ is the constant electron temperature, $n_0$ is the constant electron density and $<\varphi>=\int dz \varphi$ indicates an average along the direction of the magnetic field.  Eq. (\ref{poiss1}) corresponds, in the limit of flat density and temperature gradients, to the quasi-neutrality relation adopted in Ref. \cite{Gra06}. This relation assumes adiabatic electrons, small potential fluctuations, and accounts for a polarization term.  In this case, the expression for the operator $L$, can formally be obtained by the relation
\ben
\varphi =\left(- \Delta_{\perp} +\frac{e^2 B^2}{M T_e}(I - < >)\right)^{-1}  \circ \left( I - \int d^3 x \right) \frac{e^2 B^2}{M n_0} \int dv f,
\een
where $I$ indicates the identity operator. Note that the constant electron density is constrained by the relation $n_0=\int d^3x dv f$, expressing the equality in the number of electrons and ions in the unit cube. 

We remark that also in the presence of density and temperature gradients, which are used for instance for the study of ion temperature gradient driven turbulence, our procedure applies, with slight modifications. A detailed study of the Hamiltonian two-moment fluid model obtained by drift-kinetic models in the presence of such gradients is the subject of a forthcoming publication.

On the other hand, an example of Poisson's equation simpler than Eq. (\ref{poiss1}) and that is also covered by our procedure, corresponds to the drift-kinetic quasi-neutrality relation adopted in Ref. \cite{Mor07}, which reads
\ben  \label{poiss2}
\frac{e \varphi}{T_e}=\frac{1}{n_0} \left(\int dv f - n_0 \right).
\een
The relation (\ref{poiss2}) can be obtained from Eq. (\ref{poiss1}) by neglecting the polarization term and the flux surface average. Eq. (\ref{poiss2}) can be readily cast in the form (\ref{poiss}) as
\ben
\varphi=\left( I - \int d^3 x \right) \frac{T_e}{e n_0} \int dv f,
\een
from which it follows that in this case, the formal expression for $L$ reads $L=(I - \int d^3 x) T_e / e n_0$.
 
Assuming the above hypotheses, the drift-kinetic equation (\ref{dk}) can be written as an infinite-dimensional Hamiltonian system resulting from the Hamiltonian functional 
\beq  \label{ham}
 H(f)=\int d^3 x dv f(\bx ,v) \left(M\frac{v^2}{2}+\mathcal{B}(x)+q\frac{L}{2}\int dv' f(\bx , v') \right),
\eeq
and the noncanonical Poisson bracket
\beq  \label{pb}
\{F,G\}=\int d^3 x dv f ( [F_f , G_f]_x + [F_f , G_f]_v ),
\eeq
where the subscripts on $F$ and $G$ indicate functional derivatives.
Indeed, direct calculations show that, assuming that boundary terms vanish when integrating by parts, the equation of motion 
\beq
\frac{\pa f}{\pa t}=\{f, H\},
\eeq
yields namely the drift-kinetic equation (\ref{dk}) when the Hamiltonian (\ref{ham}) and the bracket (\ref{pb}) are used. 
We remark that (\ref{pb}) is a legitimate Poisson bracket of Lie-Poisson type \cite{Mar02,Mor98}.

We also note that the Hamiltonian (\ref{ham}) can be naturally interpreted as the total energy of the system, with its first two terms representing the kinetic energy due to the motions parallel and perpendicular to $\hat{z}$, respectively, whereas the third term accounts for the electrostatic energy.

\subsection{Poisson bracket for the kinetic moments}    \label{ssec:clof}

In order to simplify the notation, from now on, we set $M=B=1$ and $q=-1$.

Given a non-negative integer $n$, we denote with $P_n$ the kinetic moment (or simply the moment) of order $n$ of the distribution function $f$, and we define it as
\beq   \label{mom}
P_n(\bx)=\int dv v^n f(\bx ,v).
\eeq
Evolution equations for the moments of $f$ can be derived from Eq. (\ref{dk}). This leads to an infinite hierarchy of equations. As is well known, the evolution equation for the moment of order $n$, involves the moment of order $n+1$, which leads to a closure problem, in order to derive a finite system of equations.
Here we consider systems involving only the first two moments, thus implying a closure through which $P_2$ is expressed in terms of $P_0$ and $P_1$. In particular, we investigate the conditions under which the resulting closed system still possesses a Hamiltonian structure. For this purpose, we first see how the Poisson bracket (\ref{pb}) transforms when expressing it in terms of the moments and restricting its action to functionals of only $P_0$ and $P_1$. This is accomplished by first recalling that the change of variables $f \rightarrow \{P_n \}_{n \in \mathbb{N}} $ induces the following transformation:
\beq  \label{chdf}
F_f = \sum_{n \in \mathbb{N}} v^n \bar{F}_n,
\eeq
between the functional derivatives with respect to $f$ and those with respect to the moments (we indicate with $\bar{F}_n$, the functional derivative of $\bar{F}$ with respect to $P_n$). Therefore, using (\ref{mom}) and (\ref{chdf}), the bracket (\ref{pb}) transforms into
\beq  \label{pbm}
\{F,G\}=\sum_{m,n \in \mathbb{N} }\int d^3 x [P_{m+n} [F_m , G_n]_x + P_{m+n-1} (n G_n \pa_z F_m - m F_m \pa_z G_n)],
\eeq  
This operation naturally appears as the sum of two contributions: the first one, where $[ , ]_x$ appears, originates from the terms which typically account for the advection in the plane perpendicular to the magnetic field. The second contribution corresponds (apart from the additional dependence and integration with respect to $x$ and $y$) to the Lie-Poisson bracket originating from taking moments of the one-dimensional Vlasov equation \cite{Kup78,Gib81}.

If we now restrict to $F$ and $G$ functionals of $P_0$ and $P_1$ only, the operation (\ref{pbm}) reduces to
\beq  \label{bk01}
\begin{split}
&\{F,G\}=\int d^3 x \left[ P_0 [F_0 , G_0]_x + P_1 ([F_1 , G_0]_x +[F_0 , G_1]_x) + P_2 [F_1 , G_1]_x \right.\\
&\left. + P_0 (G_1 \pa_z F_0 - F_1 \pa_z G_0)+P_1 (G_1 \pa_z F_1 - F_1 \pa_z G_1 )\right].
\end{split}
\eeq
The bracket (\ref{bk01}), although bilinear, antisymmetric and satisfying the Leibniz identity, is not guaranteed to be a Poisson bracket, because truncations of Poisson brackets do not preserve the Jacobi identity, in general. This can be the case, however, if one restricts to functionals that form a sub-algebra. This does not occur for the case at hand, though. Indeed, one can see from (\ref{bk01}) that $\{ F,G\}$ depends explicitly on $P_2$. Therefore the set of functionals of $P_0$ and $P_1$ is not closed under the bracket $\{ , \}$. Remark that it is the perpendicular advection term which is responsible for the explicit dependence on $P_2$, whereas the Vlasov part of (\ref{bk01}) depends only on $P_0$ and $P_1$. We conclude then that, unlike the case of the Vlasov equation \cite{Gib08,DeG12}, for the drift-kinetic dynamics, the set of functionals of the first two moments does not form a sub-algebra.

Because two-moment fluid models are obtained assuming a closure relation that expresses $P_2$ in terms of $P_0$ and $P_1$, we ask whether a function $\calf$ exists, such that, imposing $P_2=\calf (P_0 ,P_1)$ in (\ref{bk01}), yields a Poisson bracket. In general this will not be the case because the Jacobi identity is typically not preserved by this operation. When this occurs, on the other hand, the corresponding relation $P_2=\calf (P_0 , P_1)$ is a Hamiltonian closure. For the bracket (\ref{bk01}) it turns out that, as shown in \ref{appa}, the only Hamiltonian closure (not involving differential or integral operators ) is given by
\ben \label{clos}
P_2 = \frac{P_1^2}{P_0}+\cala P_0^3,
\een
where $\cala$ is a constant. The Poisson bracket (\ref{bk01}) then becomes
\beq  \label{bkc}
\begin{split}
&\{F,G\}=\int d^3 x \left[ P_0 [F_0 , G_0]_x + P_1 ([F_1 , G_0]_x +[F_0 , G_1]_x) + \left(\frac{P_1^2}{P_0}+\cala P_0^3\right) [F_1 , G_1]_x \right.\\
&\left. + P_0 (G_1 \pa_z F_0 - F_1 \pa_z G_0)+P_1 (G_1 \pa_z F_1 - F_1 \pa_z G_1 )\right].
\end{split}
\eeq 
We remark that, this result, obtained purely by considerations based on the algebraic properties of the bracket in terms of the moments, has a clear physical interpretation. Indeed, the moments $P_0$ and $P_1$ naturally correspond to the guiding center density and the momentum, respectively of the plasma species under consideration. The second order moment, on the other hand, is related to the pressure $\mathbb{P}$ via the relation $P_2=\mathbb{P}+P_1^2 /P_0$. Therefore, the relation (\ref{clos}) corresponds to $\mathbb{P}=\cala P_0^3$. This closure relation is what one obtains for an ideal adiabatic gas possessing 3 as adiabatic index. This corresponds to gas molecules possessing one degree of freedom. In the $\cala=0$ limit one recovers the cold plasma case.

We recall that this closure in particular corresponds to an exact closure \cite{Mor07} for a water-bag distribution function of the form
\ben  \label{wb}
f(\bx ,v)=\left\{
  \begin{array}{l l}
    1/2 \sqrt{3\cala} & \quad \text{for $v_-(\bx) \leq v \leq v_+ (\bx)$ }\\
     0 & \quad \text{elsewhere},
  \end{array} \right.
\een
where the values of the velocities $v_{\pm}(\bx)$ locally determine the support of $f$. Water-bag distribution functions provide the simplest model for a kinetic distribution with a finite temperature. The latter is associated with the parameter $\cala$.

Of course, once a Poisson bracket in terms of the moments is identified, a Hamiltonian functional is required to generate the fluid dynamical equations, and different Hamiltonians lead to different dynamics. Here, it is natural to constrain the Hamiltonian functional in such a way that the resulting dynamical equations are compatible with the equations obtained taking moments of (\ref{dk}). 
Given a generic Hamiltonian functional $H(P_0 , P_1)$ and the Poisson bracket (\ref{bkc}), the resulting equations of motion are
\ben
\frac{\pa P_0}{\pa t}= [H_0 ,P_0]_x +[H_1 ,P_1]_x -\paz (P_0 H_1), \label{p0}\\
\frac{\pa P_1}{\pa t}= [H_0 , P_1]_x +\left[H_1 ,\frac{P_1^2}{P_0}+ \cala P_0^3\right]_x -P_0 \paz H_0 -\paz (P_1 H_1 )-P_1 \paz H_1.  \label{p1}  
\een
On the other hand, the equations obtained from the first two moments of (\ref{dk}) are
\ben
\frac{\pa P_0}{\pa t}=-[LP_0 , P_0]_x +[\mathcal{B}, P_0]_x -\paz P_1 , \label{p0d}\\
\frac{\pa P_1 }{\pa t}=- [L P_0 , P_1 ]_x +[\mathcal{B}, P_1]_x - \paz P_2 +P_0 \paz  L P_0. \label{p1d}
\een
Compatibility between (\ref{p0}) and (\ref{p0d}) requires $\paz (P_0 H_1) =\paz P_1$, from which we get
\beq  \label{formh}
H(P_0 , P_1)= \int d^3x \left( \frac{1}{2}\frac{P_1^2}{P_0}+\upsilon(P_0) + \lambda (x,y) \frac{P_1}{P_0} \right),
\eeq
where $\upsilon$ is in general an operator depending on $P_0$ and $\lambda$ is a function of $x$ and $y$. If one inserts the form (\ref{formh}) for the Hamiltonian into (\ref{p0}), one obtains
\beq
\frac{\pa P_0}{\pa t}=\left[\left(\int d^3x \upsilon \right)_0 , P_0 \right]_x -\left[\lambda \frac{P_1}{P_0^2} , P_0\right]_x+\left[ \frac{\lambda}{P_0} ,P_1\right]_x - \paz P_1,
\eeq
which yields (\ref{p0d}) if the Hamiltonian (\ref{formh}) specifies to
  \beq \label{h2}
H(P_0 , P_1)= \int d^3x \left[ \frac{1}{2}\frac{P_1^2}{P_0} +P_0 \left( \mathcal{B}- \frac{1}{2} L P_0 \right) + \sigma (P_0) +\lambda \frac{P_1}{P_0} \right],
\eeq
where $\sigma$ is an arbitrary function of $P_0$, $\lambda$ is constant and where we made use of the hypothesis of symmetry of the operator $L$. Inserting (\ref{h2}) into (\ref{p1}) yields
\beq \label{p1int}
\frac{\pa P_1}{\pa t}=- [L P_0 , P_1 ]_x +[\mathcal{B}, P_1]_x +[\sigma ' ,P_1]_x +\cala \left[\frac{P_1}{P_0} , P_0^3\right]_x- \paz \left(\frac{P_1^2}{P_0}\right) +P_0 \paz  L P_0 -P_0 \partial_z \sigma ',
\eeq
where the prime denotes derivative with respect to the argument of the function. Eq. (\ref{p1int}) reduces to (\ref{p1d}) namely if the closure $P_2=P_1^2 /P_0+\cala P_0^3$ is adopted and if $\sigma (P_0) =(\cala /2)P_0^3+c P_0$, where $c$ is a constant. Thus,  $P_2=P_1^2/P_0+\cala P_0^3$ appears indeed as the only possible closure for the two-moment model (\ref{p0d})-(\ref{p1d}) which is compatible with equations obtained from the Poisson bracket (\ref{bkc}).

This of course, does not exclude a priori the fact that systems obtained from other closures also admit a Hamiltonian formulation. The closure (\ref{clos}) turns out to be the only Hamiltonian closure that one obtains when imposing the closure directly in the expression for the bracket. Other closures leading to a Hamiltonian systems, if they exist, would require other Poisson brackets. 

As by-product of this procedure, we have thus also derived the explicit Hamiltonian structure for the two-moment model (\ref{p0d})-(\ref{p1d}), with the closure $P_2=P_1^2/P_0+\cala P_0^3$. Such structure consists of the Poisson bracket (\ref{bkc}) and of the Hamiltonian
\beq \label{h2m}
H=\int d^3  x \left( \frac{1}{2}\frac{P_1^2}{P_0} + P_0 \mathcal{B} - \frac{P_0 L P_0}{2} +\frac{\cala}{2}P_0^3 \right).
\eeq
Notice that in (\ref{h2m}), we have omitted the contributions associated with $\lambda (P_1/P_0)$ and $c P_0$. Indeed, direct calculations show that
\beq  \label{casim}
C_1=c \int d^3 x P_0 , \qquad C_2=\lambda \int d^3 x \frac{P_1}{P_0},
\eeq
are Casimirs for the bracket (\ref{bkc}), that is $\{C_i , F\}=0$, for $i=1,2$ and for all functionals $F(P_0, P_1)$. Consequently they do not contribute to the equations of motion and therefore their presence in the Hamiltonian is irrelevant.

We remark that, by comparing the moment Hamiltonian (\ref{h2m}) with the expression (\ref{ham}) for the original drift-kinetic Hamiltonian, one realizes that the  the former could have been obtained from the latter simply by using the definition of $P_0$ and the closure relation $P_2=P_1^2/P_0+\cala P_0^3$.  In general, however, this is not always possible, as will be the case, for instance, for the model treated in Sec. \ref{sec:df}.  

Further physical insight on the two-moment model can be obtained by attributing to $P_0$ and $P_1$ their usual meaning of density and momentum, respectively. Thus, we set $P_0=n$ and $P_1=nu$, where $n$ and $u$ are density and velocity fields, respectively. Recalling that $\varphi=L P_0$ is the electrostatic potential and that $\mathcal{B}=\mu B(1-x/l)$ is associated with an inhomogeneous magnetic field, the two-moment model with the Hamiltonian closure can be written as
\ben
\frac{\pa n}{\pa t}-q [\varphi , n]_x -\frac{\mu}{q l}\frac{\pa n}{\pa y}+\frac{\pa (nu)}{\pa z}=0, \label{mom0}\\
\frac{\pa u}{\pa t}-q[\varphi , u]_x - \frac{\mu}{q l}\frac{\pa u}{\pa y}+\frac{1}{M}\frac{\pa}{\pa z}\left(M\frac{u^2}{2}+q\varphi + \frac{3}{2}\cala n^2\right)=0, \label{mom1}
\een
where we also restored physical constants.

Eq. (\ref{mom0}) is a continuity equation expressing the advection of the guiding center density by means of the $\mathbf{E}\times\mathbf{B}$ and grad $B$ velocities and the free transport along the magnetic field. Eq. (\ref{mom1}), on the other hand, reflects the fact that velocity variations are due, again to $\mathbf{E}\times\mathbf{B}$ and grad $B$ advection in the plane perpendicular to the magnetic field, and to energy gradients along the magnetic field.

For this example, the Hamiltonian becomes
\beq  \label{hamex}
H(n,u)=\int d^3x n \left(\frac{M}{2} u^2 + \mu B(1-x/l)+q \varphi  +\frac{\cala}{2}n^2 \right),
\eeq
Thus, (\ref{hamex}) naturally expresses the total energy of the system, given by the sum of the kinetic, electrostatic and internal energies. 

In the absence of magnetic field inhomogeneity and assuming adiabatic electrons, Eqs. (\ref{mom0})-(\ref{mom1}) correspond to the mono water-bag version of the model derived in Ref. \cite{Mor07}, for which we then provide here the corresponding Hamiltonian structure. On the other hand, in the cold plasma limit $\cala=0$ and assuming a reference potential as stream function, Eqs. (\ref{mom0})-(\ref{mom1}), yield a dissipationless, slab version of the model adopted in Ref. \cite{Sch11,Gar99} to investigate Kelvin-Helmholtz instabilities at the edge of tokamak devices. 

It is interesting to note that, when expressed in terms of the variables $n$ and $u$, the Poisson bracket (\ref{bkc}), for unitary mass, takes the remarkably simpler form
\beq   \label{brsimp}
\begin{split}
& \{F , G \}= \int d^3 x \left[ n [F_n , G_n]_x + u ( [ F_u , G_n ]_x + [F_n , G_u ]_x) + 3 \cala n [F_u , G_u ]_x \right.\\
& \left. + G_u \paz F_n - F_u \paz G_n .\right]
\end{split}
\eeq
In terms of these variables, one sees that the bracket is given by the sum of a bracket of Lie-Poisson type, corresponding to the first line in Eq. (\ref{brsimp}), with another contribution involving derivatives with respect to $z$. This bracket satisfies the criterion of Ref. \cite{Tas10} to build Poisson brackets  involving $z$ derivatives, starting from a Lie-Poisson bracket defined on the $xy$ plane. This provides a further confirmation that the form (\ref{bkc}) is indeed a valid Poisson bracket. Also, we observe that $u$ and $n$, in addition to be physically relevant variables, correspond also to the normal fields suggested by the forms of the Casimirs (\ref{casim}). The Casimirs express, for this model, the conservation of the total number of particles and of the total fluid velocity and, in general, according to Ref. \cite{Tas08}, suggest a set of natural dynamical variables ($P_0$ and $P_1 / P_0$ in this case, which namely correspond to $n$ and $u$), in terms of which the Poisson bracket simplifies to a form  that makes the conservation laws more evident. Indeed, the conservation of $\int d^3 x n$ and $\int d^3 x u$ follows easily from the expression (\ref{brsimp}).


\section{Hamiltonian two-moment models from perturbed drift-kinetic equations}  \label{sec:df}


In this section we consider the distribution function as given by the sum of a background time and space independent distribution $F_{eq}(v)$, and of a time-dependent perturbation $\tf(\bx , v)$, both assumed to decay for $v \rightarrow \pm \infty$, and consider the following evolution equation for $\tf$,
\beq  \label{df}
\frac{\pa \tf}{\pa t}+[\tilde{\phi},\tf]_x -[\mathcal{B} , \tf]_x+v \frac{\pa \tf}{\pa z}+A F_{eq}'\frac{\pa \tilde{\phi}}{\pa z}=0,
\eeq
where the ``potential'' $\tilde{\phi}$ has the form
\beq  \label{Phi}
\tilde{\phi} (\bx , v)=L_0 \int dv' \tf (\bx , v') + v L_1 \int dv' v' \tf (\bx ,v'),
\eeq
with $L_0$ and $L_1$, analogously to $L$, linear symmetric operators on $L^2 (\mathcal{D})$, with coefficients independent on $v$. Note that the form (\ref{Phi}) is more general than that allowed for $\varphi$ in Sec. \ref{sec:dk}. 
Indeed, as it will be seen in the following, this greater generality will allow to treat, for instance, also some electromagnetic models.  

The equation (\ref{df}), in particular cases, reduces to some models that can be of interest when the plasma exhibits mean profiles which are weakly dependent on time and space, as in the case of tokamak turbulence. For instance, if one identifies $\tilde{\phi}$ with the electrostatic potential perturbations (and consequently considers $L_1=0$ in (\ref{Phi})), then Eq. (\ref{df}) can  be obtained from (\ref{dk}) assuming $f=F_{eq} + \tf$ and neglecting the nonlinear term involving $\partial_z \tilde{\phi} \partial_v \tf $  (this last step can be justified assuming that, in dimensionless units, $\partial_t \sim \tf / F_{eq} \sim \tilde{\phi} \sim \mathcal{B} \sim \partial_z / \partial_x \sim \partial_z / \partial_y \ll 1$).

The equation (\ref{df}), complemented by (\ref{Phi}) also admits a Hamiltonian formulation, given by the Hamiltonian
\beq  \label{hdf}
H(\tf)=\frac{1}{2}\int d^3 x dv \left[ \tf (- \tilde{\phi}+2 \mathcal{B}) - \frac{v}{A F_{eq} '}\tf^2  \right]
\eeq   
and by the Poisson bracket
\beq  \label{driftpbdf}
\{ F,G\}= \int d^3 x dv [\tf [F_{\tf} , G_{\tf}]_x +A F_{eq} ' F_{\tf} \pa_z G_{\tf}].
\eeq
The bracket (\ref{driftpbdf}) and the Hamiltonian (\ref{hdf}), in the limit $\mathcal{B}=0$, were presented in Ref. \cite{DeB01}.

Note that the Hamiltonian (\ref{hdf}) is given by the sum of a "potential energy" part, analogous to that of Eq. (\ref{ham}), with a contribution quadratic in $\tf$, which appears in the Hamiltonian for the linearized Vlasov equation \cite{Hol85,Mor94}. Similarly, the Poisson bracket (\ref{driftpbdf}) is given by the sum of a contribution analogous to the one present in (\ref{pb}) and of a term corresponding to the linearization about $F_{eq}$, of the second term of the bracket (\ref{pb}).
   
By analogy with  Sec. \ref{ssec:clof}, we define
\beq   \label{momdf}
\tilde{P}_n(\bx)=\int dv v^n \tf(\bx ,v),
\eeq
From (\ref{driftpbdf}), by adopting the functional chain rule, we obtain  the following bracket:
\beq \label{bkdf01gen}
\begin{split}
\{F,G\}= \sum_{m,n \in \mathbb{N}}\int d^3 x [ \tilde{P}_{m+n} [F_m , G_n ]_x  - A (m+n) F_m \pa_z G_n   \int dv v^{m+n-1} F_{eq} ]. 
\end{split}
\eeq
If we restrict to functionals of $\tpo$ and $\tpu$, the expression (\ref{bkdf01gen}) yields
\beq  \label{bkdf01}
\begin{split}
&\{F,G\}=\int d^3 x \left[ \tpo [F_0 , G_0]_x + \tpu ([F_1 , G_0]_x +[F_0 , G_1]_x) + \tpd [F_1 , G_1]_x \right.\\
&\left. -A n_0 (F_1\pa_z G_0 + F_0  \pa_z G_1 ) -2A M_0 F_1 \pa_z G_1\right],
\end{split}
\eeq
where
\beq
n_0=\int dv F_{eq} , \qquad M_0= \int dv v F_{eq},
\eeq
are two constants. In (\ref{bkdf01gen}) and in the following, in order not to introduce further symbols and keep a simple notation, we used the subscripts on functionals also to indicate functional derivatives with respect to $\tpo, \tpu, \tpd,..$. 
Note that, in order to obtain (\ref{bkdf01}), the property $\int dv F_{eq} '=0$, which is a consequence of the boundary conditions, has been used.
 
As is evident, also the set of functionals of $\tpo$ and $\tpu$ is not closed under the operation (\ref{bkdf01}), which depends explicitly on $\tpd$. As in Sec. \ref{ssec:clof}, we then look for functions $\tilde{\calf}(\tpo , \tpu)$ such that, when replacing $\tpd=\tilde{\calf}(\tpo , \tpu)$ in (\ref{bkdf01}), the resulting operation satisfies the Jacobi identity (bilinearity, antisymmetry and Leibniz identity, also in this case, are evidently satisfied). 

The result, whose derivation is presented in \ref{appb}, is that the only closure that preserves the Jacobi identity is
\beq  \label{hamclodf}
\tpd=a \tpo  + 2 \frac{M_0}{n_0} \tpu,
\eeq
where $a$ is an arbitrary constant. It can be verified that this result is also consistent with the criterion, derived in Ref. \cite{Tas10}, to build Poisson bracket for three-dimensional fluid models from two-dimensional Hamiltonian models.

The Hamiltonian closure (\ref{hamclodf}) can be related to the Hamiltonian closure (\ref{clos}) derived for the full drift-kinetic model. Indeed, upon recalling that  the perturbed drift-kinetic model is based on the hypothesis $f=F_{eq} + \tf$, with $\tf / F_{eq} \ll 1$,  one can apply this assumption to the Hamiltonian closure (\ref{clos}) and expand the closure relation around $P_0=n_0$, $P_1=M_0$ and $P_2={P_2}_0$, where ${P_2}_0=\int dv v^2 F_{eq}$, assuming small fluctuations of the moments. Neglecting terms quadratic in the perturbations, one then obtains
\beq  \label{isot}
\begin{split}
&\mathbb{P}=\int dv (F_{eq} + \tf) \left(v -\frac{P_1}{P_0}\right)^2 \\
&={P_2}_0-\frac{M_0^2}{n_0} + \frac{M_0^2}{n_0^2}\tpo - 2 \frac{M_0}{n_0} \tpu + \tpd + \cdots= \cala n_0^3 + 3 \cala n_0^2 \tpo + \cdots.
\end{split}
\eeq 
Eq. (\ref{isot}) implies, at the zero order,
\beq
{P_2}_0=\frac{M_0^2}{n_0}+\cala n_0^3,
\eeq
which is a condition on $F_{eq}$. At the first order, on the other hand, the Hamiltonian closure implies
\beq  \label{iso1}
\tpd=\left(3 \cala n_0^2 - \frac{M_0^2}{n_0^2}\right)\tpo + 2 \frac{M_0}{n_0} \tpu.
\eeq
The relation (\ref{iso1}) corresponds namely to the closure (\ref{hamclodf}) in the specific case $a=3 \cala n_0^2 - M_0^2 / n_0^2$. Therefore, the Hamiltonian closure (\ref{hamclodf}) of the perturbed drift-kinetic system, can be seen as the linearization of the Hamiltonian closure of the corresponding full-drift-kinetic model. The presence of the arbitrary constant $a$ reflects the presence of arbitrary constant $\cala$ of the full model, which is related to the temperature of the plasma. More in general, the relation $a=3 \cala n_0^2 - M_0^2 / n_0^2$ indicates that the constant $a$ represents the difference between the thermal energy and the parallel kinetic energy of the equilibrium state. In particular, one has $a=0$ when the plasma is cold and possesses no equilibrium flow, or, more in general, when the two energies coincide.  

Inserting the closure (\ref{hamclodf}) into (\ref{bkdf01}), one obtains the following explicit expression for the Poisson bracket 
\beq  \label{pbdf}
\begin{split}
&\{F,G\}=\int d^3 x \left[ \tpo [F_0 , G_0]_x + \tpu ([F_1 , G_0]_x +[F_0 , G_1]_x) + (a \tpo + 2 (M_0 / n_0) \tpu) [F_1 , G_1]_x \right.\\
&\left. -A n_0 (F_1\pa_z G_0 + F_0  \pa_z G_1 ) -2A M_0 F_1 \pa_z G_1\right].
\end{split}
\eeq
Considering a generic Hamiltonian $H(\tpo , \tpu)$, the Poisson bracket (\ref{pbdf}) generates the following dynamical equations
\ben  \label{p0hdf}
\frac{\pa \tpo}{\pa t}=[H_0 , \tpo]_x +[H_1 , \tpu]_x -A n_0 \pa_z H_1,
\een
\beq  \label{p1hdf}
\begin{split}
 &\frac{\pa \tpu}{\pa t}=[H_0 , \tpu]_x +a [H_1 , \tpo]_x + 2\frac{M_0}{n_0}[H_1 , \tpu]_x \\
 &  -A n_0 \pa_z H_0-2 A M_0 \pa_z H_1.
\end{split}
\eeq
On the other hand, the first two moments of (\ref{df}), upon imposing the closure (\ref{hamclodf}), evolve according to
\ben      \label{p0df}
\frac{\pa \tpo}{\pa t}=-[L_0 \tpo , \tpo]_x -[L_1 \tpu , \tpu]_x +[\mathcal{B} , \tpo]_x -\pa_z \tpu + A n_0 \pa_z L_1 \tpu,
\een
\beq  \label{p1df}
\begin{split}
&\frac{\pa \tpu}{\pa t}=-[L_0 \tpo , \tpu]_x- a [L_1 \tpu, \tpo]_x - 2 \frac{M_0}{n_0}[L_1 \tpu , \tpu]_x + [\mathcal{B} , \tpu]_x -a \pa_z \tpo\\
& - 2 \frac{M_0}{n_0} \pa_z \tpu +A n_0 \pa_z L_0 \tpo +2A M_0 \pa_z L_1 \tpu .
\end{split}
\eeq
We observe that, unlike in the case of the expression (\ref{ham}), the Hamiltonian (\ref{hdf}) cannot be expressed in terms of moments, because of the quadratic dependence on $\tf$. Nevertheless, for $A \neq 0$, equivalence between (\ref{p0hdf})-(\ref{p1hdf}) and (\ref{p0df})-(\ref{p1df}) can be obtained using, as Hamiltonian
\beq  \label{hamdf}
H(\tpo , \tpu)=\frac{1}{2} \int d^3 x \left( - \tpo L_0 \tpo + \frac{a}{A n_0} \tpo^2  + 2\tpo \mathcal{B} - \tpu L_1 \tpu +\frac{\tpu^2}{A n_0} \right).
\eeq
Thus, the Poisson bracket (\ref{pbdf}) and the Hamiltonian (\ref{hamdf}) generate the fluid equations obtained from taking the first two moments of the perturbed drift-kinetic equation (\ref{df}) and adopting the closure (\ref{hamclodf}).

Casimir invariants of the Poisson bracket (\ref{pbdf}) are given by
\beq  \label{casimred}
C_1 =\int d^3 x \tpo , \qquad C_2 =\int d^3 x \tpu.
\eeq
It is worth, in this context, considering also the two-dimensional (2D) limit of this system, assuming that $z$ is an ignorable coordinate, which is a common practice, in reduced plasma models for tokamaks, in which $z$ mimicks the toroidal coordinate. The Poisson bracket (\ref{pbdf}), in the 2D limit, reduces to
\beq  \label{pbdf2d}
\begin{split}
&\{F,G\}=\int d^2 x \left[ \tpo [F_0 , G_0]_x + \tpu ([F_1 , G_0]_x +[F_0 , G_1]_x) + (a \tpo + 2 (M_0 / n_0) \tpu) [F_1 , G_1]_x \right],
\end{split}
\eeq
which is a Poisson bracket of Lie-Poisson form. Direct calculations show that, unlike its 3D extension, the Poisson bracket (\ref{pbdf2d}) possesses two infinite families of Casimir invariants, which correspond to
\beq  \label{cas2d}
C_{\pm} = \int d^2 x \mathcal{C}_{\pm} \left[\tpu - \left( \frac{M_0}{n_0} \mp \sqrt{\frac{M_0^2}{n_0^2} + a}\right) \tpo\right],
\eeq
where $\mathcal{C}_{\pm}$ are two arbitrary functions. As will be seen in the remainder of the article, invariants of several 2D reduced fluid models for plasmas are indeed of the form (\ref{cas2d}), and can thus be interpreted as invariants of a model obtained as a two-moment hierarchy of a drift-kinetic model. In the general case, the linear combinations $\mathcal{L}_{\pm}=\tpu - \left( \frac{M_0}{n_0} \mp \sqrt{\frac{M_0^2}{n_0^2} + a}\right) \tpo$ represent an alternative set of variables, in terms of which, the model takes the form of two advection equations for $\mathcal{L}_{\pm}$. These can thus be seen as two Lagrangian invariants, formally advected by two incompressible velocity fields whose stream functions correspond to $H_{\mathcal{L}_{\pm}}$. In the degenerate case $a=M_0=0$, however, the Poisson bracket takes a semi direct product form and the two Lagrangian invariants collapse, leaving only one advected quantity in the system, namely $\tpu$. Notice also that, if one uses for $a$ the above expression obtained from the linearization of the Hamiltonian closure of the full model, then the Lagrangian invariants take the form $\mathcal{L}_{\pm}=\tpu - \left( M_0 / n_0 \mp \sqrt{ 3 \cala n_0^2 }\right) \tpo$, which is always defined since $\cala \geq 0$. From this expression one sees that the Lagrangian invariants correspond to generalized momenta given by the difference between two contributions. The first contribution $\tpu$ corresponds to the fluctuating part of the actual fluid momentum, whereas the second contribution is a momentum given by the product of the density fluctuations with a velocity depending on equilibrium quantities. This velocity, depending on whether one considers $\mathcal{L}_+$ or $\mathcal{L}_-, $ corresponds to the difference or the sum between the equilibrium mean flow and the thermal velocity.

It is interesting to remark that the Poisson bracket (\ref{brsimp}), which is the Poisson bracket for the fluid system obtained form the full drift-kinetic model in terms of the variables $n$ and $u$, is of the same type of the bracket (\ref{pbdf}), which refers to the models derived from the perturbed drift-kinetic system. More precisely, the bracket of the fluid models obtained from the full drift-kinetic system, in terms of $n$ and $u$, corresponds to  the bracket of the reduced fluid models for $\tpo=n$, $\tpu=u$, $a=3 \cala$ and $M_0=0$. Thus, the fluid models of Sec. \ref{ssec:clof}, when expressed in terms of $n$ and $u$, turn out to possess the same Poisson structure of the reduced models of Sec. \ref{sec:df} expressed in terms of the kinetic moments and in the absence of equilibrium flow. Apart from the difference in the use of the velocity instead of the momentum, the difference in the dynamical equations between the two models clearly comes from the Hamiltonian. Indeed, note that, even in terms of the variables $n$ and $u$, the Hamiltonian (\ref{hamex}) possesses cubic terms, whereas the fluid models obtained by taking moments of the perturbed drift-kinetic system are generated by Hamiltonian functionals with at most quadratic terms. 

The fact that the full and reduced fluid models, when expressed in terms of the appropriate set of variables, possess the same Poisson structure, implies of course that they possess the same Casimir functionals. Indeed, the Casimirs (\ref{casim}) are nothing but the  Casimirs (\ref{casimred}) of the reduced models, upon identifying $\tpo$ with $n$ and $\tpu$ with $P_1/ P_0$, that is, with $u$. As a consequence, also the above discussed properties of the Casimirs of the reduced models can be applied to the full fluid model. In particular, also the latter reduces to the advection of two Lagrangian invariants in the 2D limit.
For the full fluid model, the Lagrangian invariants correspond to $\mathcal{L}_{\pm}=u \pm \sqrt{3 \cala} n$. In terms of these variables, in 3D, the full model takes the symmetric form (still for $q=-1$, $B=1$ and $M=1$):
\beq
\frac{\pa \mathcal{L}_{\pm}}{\pa t}= [\bar{\varphi} , \mathcal{L}_{\pm}]_x+\frac{\pa}{\pa z}\left( \bar{\varphi}+ \frac{\mathcal{L}_{\pm}^2}{2}\right),
\eeq
where we introduced the generalized stream function $\bar{\varphi}=-\varphi + \mathcal{B}$. Thus, for finite temperature, the fluid model derived from the full drift-kinetic system can be expressed as a system of two equations of identical form, for the variables $\mathcal{L}_{\pm}$, which represent generalized velocities with thermal corrections.

\subsection{Correspondence between Hamiltonian weakly 3D fluid models and Hamiltonian drift-kinetic models}  \label{ssec:corr}

Several reduced fluid models for tokamak plasmas assume that variations of the fields along the $z$ (i.e. the ``toroidal'') coordinate, are much weaker than those along the $x$ and $y$ directions. As a consequence, the terms involving $z$-derivatives appear as linear terms in the model equations. We refer to such models as to weakly 3D models. In the non-dissipative limit, a number of 2D or weakly 3D models including reduced magnetohydrodynamics \cite{Mor84} and models for collisionless reconnection \cite{Sch94,Bor05} and interchange instability (a particular limit of the four-field model considered in Ref. \cite{Haz87} and the cold-ion limit of the model of Ref. \cite{Iza11}) have been shown to possess a Hamiltonian structure namely of the form (\ref{pbdf})-(\ref{hamdf}), where the fields $\tpo$ and $\tpu$ take different physical meanings depending on the model at hand. In particular they do not necessarily represent the density and the momentum of the plasma species. Therefore, because, due to the results of Sec. \ref{sec:df} we know that such models can be obtained by taking moments of a suitable drift-kinetic system, we can, for any weakly 3D fluid model with Hamiltonian structure  (\ref{pbdf})-(\ref{hamdf}) obtain a drift-kinetic model, with Hamiltonian structure given by (\ref{hdf})- (\ref{driftpbdf}), from which the fluid model can formally be derived. 

As a first example for this correspondence, we consider the following model for magnetic reconnection mediated by electron inertia \cite{Sch94,Bor05}:
\ben
\frac{\partial \psi_e}{\partial t}=-[\phi , \psi_e]_x +\rho_s^2 [\lap \phi , \psi]_x -\frac{\pa \phi}{\pa z}+\rho_s^2 \frac{\pa \lap \phi}{\pa z},  \label{psie}\\
\frac{\pa \lap \phi}{\pa t}=-[\phi , \lap \phi]_x +[\psi , \lap \psi]_x-\frac{\pa \lap \psi}{\pa z}, \label{vort}
\een
where $d_e$ and $\rho_s$ are constants indicating the electron skin depth and the sonic Larmor radius, respectively, and $\psi_e$ is the toroidal electron canonical momentum, defined as $\psi_e=\psi - d_e^2 \lap \psi$. In (\ref{psie})-(\ref{vort}), $\psi$ and $\phi$ indicate the magnetic flux function and the electrostatic potential, respectively.

This model is known \cite{Sch94} to possess a Hamiltonian structure with Hamiltonian given by
\beq   \label{hamrec}
H = \frac{1}{2} \int d^3 x \left( - \phi \lap \phi + \rho_s^2 (\lap \phi)^2 -\frac{\psi_e}{d_e^2} (I - d_e^2 \lap)^{-1}\psi_e + \frac{\psi_e^2}{d_e^2}   \right),
\eeq
and Poisson bracket
\beq  \label{brrec}
\begin{split}
&\{F,G\}= \int d^3 x \left[ \lap \phi [F_{\lap \phi} , G_{\lap \phi}]_x + \psi_e ( [ F_{\psi_e} , G_{\lap \phi}]_x + [F_{\lap \phi} , G_{\psi_e}]_x) + \rho_s^2 d_e^2 \lap \phi [F_{\psi_e} , G_{\psi_e}]_x  \right.\\
& \left. - F_{\psi_e} \frac{ \pa G_{\lap \phi}}{\pa z} - F_{\lap \phi} \frac{\pa G_{\psi_e}}{\pa z} \right].  
\end{split}
\eeq
The Hamiltonian (\ref{hamrec}) and the bracket (\ref{brrec}) belong to the class (\ref{hamdf}) and (\ref{pbdf}), respectively, via the following identification:
\ben
\tpo=\lap \phi, \qquad \tpu=-\frac{\psi_e}{d_e^2}, \qquad a=\frac{\rho_s^2}{d_e^2}, \qquad \mathcal{B}=0,\\
M_0=0, \qquad A  = \frac{1}{d_e^2 n_0}, \qquad L_0=\Delta^{-1} ,\qquad L_1=d_e^2 (I - d_e^2 \lap)^{-1}.
\een
Consequently, from (\ref{df})-(\ref{Phi}), we can infer that the reconnection model (\ref{psie})-(\ref{vort}) can be derived by taking the first two moments of the drift-kinetic model \cite{DeB01}
\beq  \label{dfrec}
\begin{split}
&\frac{\pa \tf}{\pa t}+[\tilde{\phi},f]_x +v \frac{\pa \tf}{\pa z}+A F_{eq} '\frac{\pa \tilde{\phi}}{\pa z}=0, \\
&\tilde{\phi} (\bx ,v)=\lap^{-1} \int dv' \tf (\bx , v') + d_e^2 v (I - d_e^2 \lap)^{-1} \int dv' v' \tf (\bx ,v'),
\end{split}
\eeq
closing the hierarchy by imposing
\beq
\tpd=\frac{\rho_s^2}{d_e^2} \tpo,
\eeq
and then finally applying the above identifications $\tpo=\lap\phi$, $\tpu=-\psi_e / d_e^2$ and $A n_0=1/ d_e^2$.  In a similar way, weakly 3D reduced magnetohydrodynamics can be obtained, when the presence of $\rho_s$ and $d_e$ is suppressed.

We remark that, according to (\ref{cas2d}), the Lagrangian invariants of the 2D limit of the model are given by
\beq
\mathcal{L}_{\pm}=-\frac{1}{d_e^2}\left(\psi_e \mp \rho_s d_e \lap \phi\right).
\eeq
The role of such invariants for magnetic reconnection, and their analogy with Lagrangian invariants in drift kinetic system have been investigated in Refs. \cite{Caf98,Lis04,Peg05,Peg05b}.

As a second example of identification between Hamiltonian structures of weakly 3D fluid models and perturbed drift-kinetic models, we consider the following system:
\ben
\frac{\pa n_e }{\pa t}=-[\phi , n_e]_x + 2 v_d \frac{\pa n_e }{\pa y}-\frac{\pa u_e}{\pa z},  \label{ele1}\\
\epsilon \frac{\pa u_e}{\pa t}=-\epsilon [\phi , u_e]_x +2 \epsilon v_d \frac{\pa u_e }{\pa y} -\frac{\pa n_e}{\pa z} +\frac{\pa \phi}{\pa z}. \label{ele2}
\een
In (\ref{ele1})-(\ref{ele2})  $n_{e}$ and $u_{e}$, in normalized form, indicate the electron density and momentum fluctuations, $\phi=\lap^{-1}n_e $ is the electrostatic potential, whereas the constants $\epsilon$ and $v_d$ indicate the electron to ion mass ratio and the gradient-$B$ drift velocity, respectively. This model corresponds to the cold ion electrostatic limit of the model of Ref. \cite{Wae12}, where the ion fluid is assumed to be static and incompressible.

Eqs. (\ref{ele1})-(\ref{ele2}) form a Hamiltonian system with Hamiltonian
\beq  \label{hamele}
\begin{split}
&H=\frac{1}{2} \int d^3 x \left( - n_e  \lap^{-1} n_e  + n_e^2  + \epsilon u_e^2   + 4  v_d x n_e \right)
\end{split}
\eeq
and Poisson bracket
\beq  \label{brele}
\begin{split}
& \{ F,G \}= \int d^3 x \left( n_e [F_{n_e} , G_{n_e}]_x + u_e ( [ F_{u_e} , G_{n_e}]_x + [F_{n_e} , G_{u_e}]_x ) \right. \\
& \left. +\frac{n_e}{\epsilon} [ F_{u_e} , G_{u_e}]_x -\frac{1}{\epsilon} ( F_{u_e} \pa_z G_{n_e} + F_{n_e} \pa_z G_{u_e}) \right).
\end{split}
\eeq
We can easily see that the bracket (\ref{brele}) is of the form (\ref{pbdf}). The identification required to build the parent drift-kinetic model is then the following:
\ben \nonumber
\tilde{P}_{0} = n_{e}, \qquad \tilde{P}_{1} = u_{e}, \qquad A=\frac{1}{\epsilon n_{0}},  \qquad \mathcal{B}=2 v_d x, \\
a=\frac{1}{\epsilon}, \qquad M_0=0, \qquad L_0=\lap^{-1}, \qquad L_1=0 \nonumber
\een
Consequently, the Hamiltonian structure for the corresponding perturbed drift-kinetic system is given by   
\beq  \label{hamdfele}
H( \tf)=\frac{1}{2} \int d^3 x dv \left( -\phi \tf  - \epsilon n_{0}\frac{v}{{F_{eq}} '} \tf^2  + 4  v_d x \tf \right) 
\eeq
\beq  \label{brdfele}
\begin{split}
\{F, G\}= \int d^3 x dv \left( \tf [F_{\tf} , G_{\tf} ]_x +\frac{{F_{eq}} '}{\epsilon n_{0}} F_{\tf } \pa_z G_{\tf}  \right)
\end{split}
\eeq
The corresponding dynamical equation is then given by
\ben
\frac{\pa \tf}{\pa t}=-[\phi , \tf]_x + 2 v_d \frac{\pa \tf}{\pa y}-\frac{{F_{eq}} '}{\epsilon n_{0}} \frac{\pa \phi}{\pa z} - v \frac{\pa \tf}{\pa z} .  \label{dfele1}
\een
The fluid model (\ref{ele1})-(\ref{ele2}) can indeed be obtained by taking the first two moments of (\ref{dfele1}) and imposing the closure
\beq
\tilde{P}_{2}= \frac{\tilde{P}_{0}}{\epsilon}.
\eeq
We anticipate that the above described procedure can be extended to models involving more than one species. For instance, upon restoring the ion dynamics in the model (\ref{ele1})-(\ref{ele2}), one obtains the system 
\ben
\frac{\pa n_e }{\pa t}=-[\phi , n_e]_x + 2 v_d \frac{\pa n_e }{\pa y}-\frac{\pa u_e}{\pa z},  \label{elei1}\\
\epsilon \frac{\pa u_e}{\pa t}=-\epsilon [\phi , u_e]_x +2 \epsilon v_d \frac{\pa u_e }{\pa y} -\frac{\pa n_e}{\pa z} +\frac{\pa \phi}{\pa z}, \\
\frac{\pa n_i}{\pa t}= -[\phi , n_i]_x - 2 v_d \tau \frac{\pa n_i }{\pa y} -\frac{\pa u_i}{\pa z},\\
\frac{\pa u_i }{\pa t}=-[\phi , u_i ]_x -2 v_d \tau \frac{\pa u_i}{\pa y}- \tau \frac{\pa n_i }{\pa z} -\frac{\pa \phi}{\pa z}, \label{elei4}
\een
where the subscript $i$ refers to ion quantities, $\tau$ is the constant ratio between ion and electron temperature and  $\phi=\lap^{-1}(n_e - n_i)$. Eqs. (\ref{elei1})-(\ref{elei4}) form a Hamiltonian system with Hamiltonian
\beq  \label{hamele2}
\begin{split}
&H=\frac{1}{2} \int d^3 x \left[ - (n_e - n_i) \lap^{-1}(n_e - n_i) + n_e^2 + \tau n_i^2 + \epsilon u_e^2 + u_i^2 \right.\\
 & \left. + 4  v_d x (n_e + \tau n_i)\right] 
\end{split}
\eeq
and Poisson bracket
\beq  \label{brele2}
\begin{split}
& \{ F,G \}= \int d^3 x \left( n_e [F_{n_e} , G_{n_e}]_x + u_e ( [ F_{u_e} , G_{n_e}]_x + [F_{n_e} , G_{u_e}]_x ) \right. \\
& \left. +\frac{n_e}{\epsilon} [ F_{u_e} , G_{u_e}]_x -\frac{1}{\epsilon} ( F_{u_e} \pa_z G_{n_e} + F_{n_e} \pa_z G_{u_e}) \right.\\
& \left. - n_i [F_{n_i} , G_{n_i}]_x - u_i ( [ F_{u_i} , G_{n_i}]_x + [F_{n_i} , G_{u_i}]_x)  \right. \\
& \left. - \tau n_i [ F_{u_i} , G_{u_i}]_x - ( F_{u_i} \pa_z G_{n_i} + F_{n_i} \pa_z G_{u_i})\right).
\end{split}
\eeq
The corresponding parent drift-kinetic model is given by
\ben
\frac{\pa \tf_e}{\pa t}=-[\phi , \tf_e]_x + 2 v_d \frac{\pa \tf_e}{\pa y}-\frac{{F_{eq}}_e '}{\epsilon n_{e0}} \frac{\pa \phi}{\pa z} - v \frac{\pa \tf_e}{\pa z} ,  \label{dfele12}\\
\frac{\pa \tf_i}{\pa t}=-[\phi , \tf_i]_x - 2 \tau v_d \frac{\pa \tf_i}{\pa y}+\frac{{F_{eq}}_i '}{ n_{i0}} \frac{\pa \phi}{\pa z} - v \frac{\pa \tf_i}{\pa z}, \label{dfele22}
\een
where $\phi=\lap^{-1}\int dv( \tf_e - \tf_i)$. The  Hamiltonian structure of the model is given by
\beq  \label{hamdfele2}
H(\tf_i , \tf_e)=\frac{1}{2} \int d^3 x dv \left( \phi (\tf_i - \tf_e) - \epsilon n_{e0}\frac{v}{{F_{eq}}_e '} \tf_e^2 -  n_{i0}\frac{v}{{F_{eq}}_i '} \tf_i^2 + 4  v_d x (\tf_e + \tau \tf_i)\right) 
\eeq
\beq  \label{brdfele2}
\begin{split}
\{F, G\}= \int d^3 x dv \left( \tf_e [F_{\tf_e} , G_{\tf_e} ]_x +\frac{{F_{eq}}_e '}{\epsilon n_{e0}} F_{\tf_e } \pa_z G_{\tf_e}  - \tf_i [F_{\tf_i} , G_{\tf_i} ]_x +\frac{{F_{eq}}_i '}{ n_{i0}} F_{\tf_i } \pa_z G_{\tf_i} \right).
\end{split}
\eeq
The fluid model (\ref{elei1})-(\ref{elei4}) can indeed be obtained by taking the first two moments of (\ref{dfele12})-(\ref{dfele22}) and imposing the closure
\beq
\tilde{P}_{e_2}= \frac{\tilde{P}_{e_0}}{\epsilon}, \qquad \tilde{P}_{i_2} = \tau \tilde{P}_{i_0},
\eeq
with the identifications $\tilde{P}_{i,e_0}=n_{i,e}$ and $\tilde{P}_{i,e_1}=u_{i,e}$.

\section{Conclusions}  \label{sec:concl}

We have pointed out that, unlike what happens for Vlasov systems, for Hamiltonian drift-kinetic systems, the set of functionals of the first two moments does not form a sub-algebra with respect to the Poisson bracket of the drift-kinetic equation expressed in terms of the moments. We investigated then under what conditions the bracket obtained by imposing a closure relation between the second order moment and the two lowest order moments, is a Poisson bracket. The analysis has been carried out in two cases : a full drift-kinetic model and a perturbed drift-kinetic model, in which a separation between an equilibrium distribution function and a perturbation is introduced. In the former case the constraint of the Jacobi identity dictates that the only closure leading to a Poisson bracket  is given by $P_2=P_1^2 / P_0 + \cala P_0^3$. This corresponds to the equation of state for an ideal adiabatic gas with molecules possessing one degree of freedom. In the limit $\cala=0$, one retrieves the cold plasma closure as a particular case. As remarked in \ref{appa}, the isothermal closure $P_2=P_1^2 / P_0 + \cala P_0^3$, on the other hand, is not a Hamiltonian closure.
  In the case of the perturbed drift-kinetic system, the Hamiltonian closure  turns out to be   $\tilde{P}_2 = a \tilde{P}_0 + 2 u_0 \tilde{P}_1$, where $a$ is an arbitrary constant and $u_0$ is the mean velocity associated with the equilibrium distribution function ( in this expression the particle mass $M$ is equal to unity). This Hamiltonian closure can be seen as the linearization of the corresponding Hamiltonian closure for the full drift-kinetic model about an equilibrium density $n_0$ and an equilibrium momentum $M_0$. The constant $a$, then takes the form $a=3 \cala n_0^2 - u_0^2$, expressing the difference between equilibrium thermal energy and parallel kinetic energy. In this context, one can also interpret the Lagrangian invariants $\mathcal{L}_{\pm}$ of the corresponding 2D reduced fluid model, as generalized momenta involving the thermal velocity and the equilibrium parallel flow.
  
  We have also shown that, when expressed in terms of the normal fields suggested by the presence of the Casimir invariants, that is $n$ and $u$, the Poisson bracket for the full fluid model takes a simpler form, which actually is of the form of the Poisson bracket for the reduced fluid models, when the latter are expressed in terms of the kinetic moments. Therefore, the full and reduced fluid models when expressed in terms of the appropriate variables, possess the same Poisson structure. This reflects also in their Casimir invariants and in particular in the possibility to write also the full fluid model in terms of  dynamical varibles $\mathcal{L}_{\pm}$ that become Lagrangian invariants in the 2D limit. For the full fluid model the variables $\mathcal{L}_{\pm}$ are linear combinations of plasma mean velocity and thermal speed and in terms of them, the system takes a remarkably symmetric form. 
  
  Our analysis, indicates that the natural constraint of preserving a Hamiltonian structure when taking moments of a Hamiltonian drift-kinetic model, selects some particular closures, which turn out to have a direct physical interpretation.  On the other hand, however, our analysis, of course does not imply that all the two-moment   Hamiltonian models obtained from taking moments of drift-kinetic systems must have one of the two above mentioned closures. A simple counterexample in this respect, can already be obtained in the case of the one-moment model
\beq   \label{cex}
\frac{\partial \tilde{P}_0}{\partial t}+[L_0 \tilde{P}_0, \tilde{P}_0]_x+\frac{\partial }{\partial z}(L_0 \tilde{P}_0 + W(\tilde{P}_0)),
\eeq
which is obtained by taking the zero order moment of Eq. (\ref{df}), in the limit $\mathcal{B}=0$, and imposing the closure $\tilde{P}_1 = L_0 \tilde{P}_0 + W(\tilde{P}_0 )$, where $W$ is a smooth function. This closure does not belong to the class of closures that we considered. Nevertheless, Eq. (\ref{cex}) is a Hamiltonian system, with Poisson bracket and Hamiltonian corresponding to $\{F,G\}=\int d^3 x (\tilde{P}_0 [F_0 , G_0]_x -F_0 \partial_z G_0)$ and $H(\tilde{P}_0)=\int d^3 x [(1/2)\tilde{P}_0 L_0 \tilde{P}_0 + \mathcal{W} (\tilde{P}_0)]$, respectively, with $\mathcal{W}$ such that $\mathcal{W} '(\tilde{P}_0) =W(\tilde{P}_0)$.

Our procedure, on the other hand, is based on imposing the closure relation directly into the bracket expressed in terms of the moments and truncated. Because of the fragility of the Jacobi identity, this operation, apart from the case of the Hamiltonian closures that we found, does not lead to Poisson brackets.

Ways alternatives to ours to derive Hamiltonian fluid models might be sought for. On one hand, indeed, we excluded the presence of differential operators in our closure relation, which in principle, could lead to further Hamiltonian closures.  Also, one might imagine in principle a Hamiltonian fluid model whose bracket does not derive from the bilinear form obtained by taking the moments of the parent drift-kinetic bracket.
We presume that, in such cases, however, the Hamiltonian structure has to be derived a posteriori, or by imposing the closure relation through procedures, such as Dirac's theory of constraints \cite{Dir50}, that automatically preserve the Hamiltonian structure. On the other hand, we believe that our procedure still covers a quite broad class of cases of interest by following a systematic procedure. 
 
 In the case of the perturbed drift-kinetic systems we also proposed a procedure to recover a Hamiltonian parent drift-kinetic model, from a given Hamiltonian fluid model. More precisely, if the Hamiltonian fluid model is characterized by a Hamiltonian functional and a Poisson bracket of the form (\ref{hamdf})  and (\ref{pbdf}), respectively, then it means that it can be derived from a Hamiltonian drift-kinetic system with Hamiltonian and Poisson bracket corresponding to (\ref{hdf}) and (\ref{driftpbdf}).
 Hamiltonian structures of  the type (\ref{hamdf})- (\ref{pbdf}) are not rare for fluid models of interest for instance, for tokamak plasmas, in which the presence of a strong magnetic field along the toroidal direction makes the assumption of "weak" dependence on the $z$ coordinate, reasonable. We showed explicitly, indeed, how, from two weakly-3D fluid models for plasmas, the corresponding parent drift-kinetic models can be derived. As anticipated at the end of Sec. \ref{sec:concl}, this procedure can be extended to systems involving more than one plasma species, and we provided an explicit example of this by retrieving the Hamiltonian drift-kinetic model of a four-field compressible electrostatic fluid model. 
 
Our analysis, however, clearly presents limitations and we believe it would be appropriate to extend it in different directions. The first evident restriction concerns not having considered moments with respect to the $\mu$ coordinate. This clearly excludes the analysis of important closures, such as those involving pressure anisotropies. Also, for simplicity, the hypothesis of a straight magnetic guide field has been made. This excludes curvature drifts, which would complexify the problem, by introducing a quadratic dependence on $v$ in the drift-kinetic equation (see, e.g. Ref. \cite{Nis00}). At the level of the equations motion, this implies that the evolution equation for the moment of order $n$, with respect to the parallel velocity, would depend on moments of order $n+1$ and $n+2$. In terms of the Hamiltonian structure, this would reflect into a generalization of the Poisson brackets (\ref{pb}) and (\ref{pbm}), and consequently of the definitions of Hamiltonian closures.

Finally, an obvious important limitation, concerns the restriction to the first two moments. Although, as already mentioned, this already permits to derive a number of significant reduced models, it prevents an accurate description of certain phenomena. For instance, the derivation of energy-conserving, and in particular Hamiltonian fluid models accounting for the evolution of quantities related to higher order moments, such as parallel and perpendicular temperatures, as well as heat fluxes, is important for the description of turbulence in both astrophysical and fusion plasmas \cite{Sco10,Sny01,Gos05}.

The extension of the present analysis in the above directions as well as the inclusion of further physical ingredients such as gyrokinetic corrections and density gradients is under development and will be the subject of forthcoming publications.

\section*{Acknowledgments}

The author acknowledges useful discussions with Phil Morrison, Cesare Tronci, George  Throumoulopoulos, Natalia Tronko and with the Nonlinear Dynamics Team of the Centre de Physique Th\'eorique. 
This work was supported by the European Community under the contracts of
Association between EURATOM, CEA, and the French Research Federation for
fusion studies. The views and opinions expressed herein do not necessarily
reflect those of the European Commission. Financial support was also
received from the Agence Nationale de la Recherche (ANR GYPSI n. 2010 BLAN
941 03) and from the CNRS through the PEPS project GEOPLASMA.


\appendix

\section{Derivation of the Hamiltonian closure for the drift-kinetic bracket}
 \label{appa}

In this Appendix we show that $P_2=P_1^2/P_0$ is the only possible closure of the type $P_2=\calf (P_0,P_1)$ , through which the bracket (\ref{bk01}) satisfies the Jacobi identity (assuming that $\calf$ does not involve integral or differential operators).

First, we perform the replacement $P_2=\calf$ and write the bracket (\ref{bk01}) as the sum of two contributions:
\beq  \label{brf}
\{F,G\}=\{F,G\}_c + \{F,G\}_h,
\eeq
where
\ben \nonumber
\{F,G\}_c=\int d^3 x \left[ P_0 [F_0 , G_0]_x + P_1 ([F_1 , G_0]_x +[F_0 , G_1]_x) + \right.\\
\left. + P_0 (G_1 \pa_z F_0 - F_1 \pa_z G_0)+P_1 (G_1 \pa_z F_1 - F_1 \pa_z G_1 )\right], \nonumber \\
\{F,G\}_h=\int d^3 x \calf [F_1 , G_1]_x. \nonumber
\een
The Jacobi identity for (\ref{brf}) then reads as
\beq \label{jac}
\begin{split}
& \{ \{ F,G\}_c,H\}_c+\{ \{ F,G\}_c,H\}_h \\
& +\{ \{ F,G\}_h,H\}_c+\{ \{ F,G\}_h,H\}_h+\circlearrowleft=0, \quad \forall F,G,H,
\end{split}
\eeq
where $\cir$ indicates the additional terms obtained by cyclic permutations.
Then, we decompose $\{ , \}_c$ in its turn, as
\beq
\{F ,G \}_c=\{F ,G\}_x+\{F, G\}_V,
\eeq
where
\ben
\{F,G\}_x=\int d^3 x \left[ P_0 [F_0 , G_0]_x + P_1 ([F_1 , G_0]_x +[F_0 , G_1]_x)\right],\\
\{F,G\}_V=\int d^3 x \left[P_0 (G_1 \pa_z F_0 - F_1 \pa_z G_0)+P_1 (G_1 \pa_z F_1 - F_1 \pa_z G_1 )\right].
\een
We remark that $\{ , \}_x$ and $\{ , \}_V$ independently satisfy the Jacobi identity. Indeed, $\{ , \}_x$ is a Poisson bracket with a semidirect product structure and it appears, for instance in the Hamiltonian formulation of reduced magnetohydrodynamics \cite{Mor84}. The bracket $\{ , \}_V$ on the other hand, is obtained from the Poisson bracket for the first two moments of the Vlasov equation \cite{Gib81,Kup78}, with an additional dependence on the $x$ and $y$ coordinates, which is irrelevant for the Jacobi identity. Consequently, we can write
\ben  \label{jacxv}
\{ \{ F,G\}_c,H\}_c + \cir =\{ \{ F,G\}_x,H\}_V+\{ \{ F,G\}_V,H\}_x+\circlearrowleft.
\een
In order to determine (\ref{jacxv}) explicitly, we first observe that
\ben 
(\{ F,G\}_x)_0=[F_0 , G_0]_x +\text{s.v.t.},  \label{df1}\\
(\{ F,G\}_x)_1=[F_1 , G_0]_x + [F_0 , G_1]_x +\text{s.v.t.},  \label{df2}  \\
(\{ F,G\}_V)_0=G_1 \paz F_0 - F_1 \paz G_0  +\text{s.v.t.},  \label{df3} \\
 (\{ F,G\}_V)_1=G_1 \paz F_1 - F_1 \paz G_1  +\text{s.v.t.}, \label{df4}
\een
where $\text{s.v.t.}$ means 'second variation terms'. Terms involving second derivatives have not been written explicitly because, by virtue of the antisymmetry they automatically vanish when computing $\{ \{ F,G\}_c,H\}_c + \cir$.

Making use of (\ref{df1})-(\ref{df2}) and of the Leibniz rule, we can then write
\beq
\begin{split}
&\{ \{ F,G\}_c,H\}_c + \cir=\{ \{ F,G\}_x,H\}_V+\{ \{ F,G\}_V,H\}_x+\circlearrowleft\\
&=\int d^3 x [ P_0 G_1 [\pa_z F_0 , H_0]_x +P_0 \pa_z F_0 [G_1 ,H_0]_x - P_0 F_1 [\pa_z G_0 , H_0]_x-P_0 \pa_z G_0 [F_1 ,H_0]_x\\
& +P_1 G_1 [\pa_z F_1 ,H_0]_x +P_1 \pa_z F_1 [G_1 , H_0]_x -P_1 F_1 [\pa_z G_1 , H_0]_x - P_1 \pa_z G_1 [F_1 , H_0]_x  \\
&+P_1 G_1 [\pa_z F_0 , H_1 ]_x +P_1 \pa_z F_0 [G_1 , H_1 ]_x -P_1 F_1 [\pa_z G_0 ,H_1 ]_x -P_1 \pa_z G_0 [F_1 ,H_1]_x + \cir.
\end{split}
\eeq
Using the identity $\pa_z [f,g]_x = [\pa_z f , g]_x + [f , \pa_z g]_x$, this expression reduces to
\ben  \label{cc}
\{ \{ F,G\}_c,H\}_c + \cir = 2 \int d^3 x P_1 \pa_z F_0 [G_1 , H_1 ]_x +\cir,
\een
which clearly does not vanish for arbitrary $F$, $G$ and $H$. Thus, we can conclude that $\{ , \}_c$ is not a Poisson bracket. In particular, this implies that $P_2=0$ is not a Hamiltonian closure.

Making use of 
\ben
(\{F, G\}_h)_0 =\pao \calf [F_1 ,G_1 ]_x +\text{s.v.t.},\\
(\{F, G\}_h)_1 =\pau \calf [F_1 ,G_1 ]_x +\text{s.v.t.},
\een
(where $\pao$ and $\pau$ indicate the partial derivatives with respect to $P_0$ and $P_1$) we can evaluate the further contributions in (\ref{jac}).
The last term on the left-hand side of (\ref{jac}) gives
\beq  \label{hh}
\begin{split}  
& \{ \{ F,G\}_h,H\}_h+\circlearrowleft=\\
& \int d^3 x \calf [ \pau \calf [F_1 , G_1 ]_x , H_1 ]_x +\cir=\\
& \int d^3 x [\calf \pau \calf  [[F_1 , G_1 ]_x , H_1 ]_x + \calf [F_1 , G_1 ]_x [\pau \calf , H_1]_x ]+\cir=\\
& \int   d^3 x \calf [F_1 , G_1 ]_x [\pau \calf , H_1]_x +\cir,
\end{split}
\eeq
where in the last step we made use of the Jacobi identity for the bracket $[ , ]_x$. 

The mixed contributions in (\ref{jac}), on the other hand, they give
\beq   \label{ch}
\begin{split}
& \{\{F,G\}_c , H\}_h + \cir=\int d^3 x \calf \left[[F_1 , G_0 ]_x + [F_0 ,G_1]_x \right. \\
& \left. +G_1 \paz F_1 - F_1 \paz G_1 , H_1 \right]_x +\cir,
\end{split}
\eeq
and
\beq  \label{hc}
\begin{split}
&\{\{F,G\}_h , H\}_c +\cir  \\
&=\int d^3 x \left[ P_0 [ \pao \calf [F_1 , G_1]_x , H_0]_x + P_1 ([ \pau \calf [F_1 , G_1 ]_x ,H_0]_x +[\pao \calf [F_1 , G_1]_x , H_1]_x) \right.  \\
&\left. +P_1(H_1 \paz ( \pau \calf [F_1 ,G_1]_x )- \pau \calf [F_1 , G_1]_x \paz H_1)\right.  \\
&\left. +P_0 (H_1 \paz ( \pao \calf [F_1 , G_1]_x)- \pau \calf [F_1 , G_1]_x \paz H_0 )\right]+\cir,  
\end{split}
\eeq
respectively.

In order for the Jacobi identity to be satisfied, the sum of (\ref{cc}), (\ref{hh}), (\ref{ch}) and (\ref{hc}) must be zero for arbitrary $F$, $G$ and $H$. At this point we observe that, namely because of the arbitrariness of the functionals and of the independence of $\calf$ on derivatives, the only way the contribution (\ref{cc}) can be cancelled, is by means of the last terms in Eq. (\ref{hc}), which possess the same dependence on the functional and partial derivatives. Considering the appropriate permutation, this cancellation occurs if and only if
\beq  \label{canc}
2 P_1 \paz H_0 [F_1 ,G_1]_x= P_0 \pau \calf \paz H_0 [F_1 , G_1]_x,
\eeq
and analogously for the terms obtained by cyclic permutations. From (\ref{canc}) then we get
\beq
\calf (P_0 ,P_1) = \frac{P_1^2}{P_0}+\theta (P_0),
\eeq
where $\theta$ is an arbitrary function.

We determine now under what conditions $\{ , \}$ satisfies the Jacobi identity with this form for $\calf$. It is practical to treat the contributions from $P_1^2 /P_0$ and from $\theta$, separately. 

Thus, if we set $\theta=0$ we get
\beq
\begin{split}
& \{\{F,G\},H\}+ \cir \stackrel{\theta=0}{=}\\
& \stackrel{\theta=0}{=}\int d^3 x \left[ 2 P_1  \paz F_0 [G_1 , H_1]_x +\frac{P_1^2}{P_0}[F_1 , G_1]_x \left[2\frac{P_1}{P_0} ,H_1\right]_x \right.\\
&\left.  + \frac{P_1^2}{P_0}([[F_1 , G_0]_x ,H_1]_x + [[F_0 , G_1]_x ,H_1]_x +[G_1 \paz F_1 , H_1]_x - [F_1 \paz G_1 , H_1]_x) \right.\\
&\left. - \frac{P_1^2}{P_0}[[F_1 , G_1 ]_x ,H_0]_x -P_0 [F_1 , G_1]_x \left[\frac{P_1^2}{P_0^2} , H_0 \right]_x+ 2 \frac{P_1^2}{P_0}[[ F_1 , G_1 ]_x ,H_0]_x +P_1 [F_1 , G_1 ]_x  \left[2\frac{P_1}{P_0} ,H_1\right]_x  \right.\\
&\left. - \frac{P_1^3}{P_0^2} [ [F_1 , G_1 ]_x ,H_1 ]_x  -P_1 [F_1 , G_1 ]_x \left[\frac{P_1^2}{P_0^2} ,H_1 \right]_x - P_0 H_1 \paz \left( \frac{P_1^2}{P_0^2} \right) [F_1 , G_1 ]_x - \frac{P_1^2}{P_0} H_1 \paz [F_1 , G_1 ]_x \right.\\
& \left. - 2 P_1 [F_1 , G_1 ]_x \paz H_0 +2 P_1 H_1 \paz \left( \frac{P_1}{P_0}\right) [F_1 , G_1 ]_x \right.\\
& \left. + 2  \frac{P_1^2}{P_0} H_1 \paz [F_1 , G_1 ]_x -2\frac{P_1^2}{P_0} \paz H_1 [F_1 , G_1]_x \right] + \cir\\
&= \int d^3 x \left[  \frac{P_1^2}{P_0} ( [[F_1 , G_0]_x , H_1]_x + [[ F_0, G_1 ]_x , H_1]_x + [[F_1 , G_1 ]_x , H_0]_x ) \right. \\
& \left. + G_1 [ \paz F_1 , H_1 ]_x + \paz F_1 [G_1 , H_1 ]_x -F_1 [ \paz G ,H_1 ]_x - \paz G_1 [F_1 , H_1 ]_x + H_1 [ \paz F_1 , G_1 ]_x \right.\\
&\left. + H_1 [F_1 , \paz G_1 ]_x - 2 \paz H_1 [F_1 , G_1]_x - \frac{P_1^3}{P_0^2} [[F_1 , G_1]_x , H_1 ]_x  \right] + \cir =0,
\end{split}
\eeq
where in the last step the coefficients of $P_1^2 /P_0$ and $P_1^3 / P_0^2$ vanish because of the Jacobi identity for $[ , ]_x$. Thus, we have shown that $\calf=P_1^2 / P_0$ yields a Hamiltonian closure. If we now consider also the contributions from $\theta$ we get
\beq  \label{the}
\begin{split}
& \{\{F,G\},H\}+ \cir =\\
& = \int d^3 x \left[ \theta ([[F_1 , G_0]_x , H_1 ]_x + [[F_0 , G_1 ]_x , H_1 ]_x +[G_1 \paz F_1 , H_1 ]_x -[ F_1 \paz G_1 , H_1 ]_x ) \right. \\
&\left.  + P_0 \theta ' [[F_1 , G_1 ]_x ,H_0 ]_x +P_0 [F_1 , G_1 ]_x [\theta ' , H_0]_x + P_1 \theta ' [[F_1 , G_1 ]_x , H_1 ]_x + P_1 [ F_1 , G_1 ]_x [\theta ' , H_1 ]_x \right.\\
& \left. + P_0 H_1 \paz \theta ' [F_1 , G_1 ]_x  + P_0 H_1 \theta ' [\paz F_1 , G_1 ]_x +P_0 H_1 \theta ' [F_1 , \paz G_1 ]_x + \theta [F_1 , G_1]_x \left[2\frac{P_1}{P_0} ,H_1\right]_x\right] + \cir \\
& =\int d^3 x \left[ (- \theta  + P_0 \theta ')[[H_1 , F_1 ]_x , G_0]_x +P_0 [F_1 , G_1 ]_x [\theta ' ,H_0]_x  +P_1 \theta' [[F_1 , G_1 ]_x ,H_1]_x + P_1 [F_1 , G_1 ]_x [\theta ' ,H_1 ]_x \right.\\
& \left. +P_0( H_1 \paz \theta ' [F_1 , G_1 ]_x + H_1 \theta ' [\paz F_1 ,G_1 ]_x + H_1 \theta ' [F_1 , \paz G_1 ]_x )\right. \\
& \left. + \theta ( G_1 [\paz F_1 ,H_1 ]_x +\paz F_1 [G_1 , H_1 ]_x -F_1 [\paz G_1 ,H_1 ]_x -\paz G_1 [F_1 , H_1 ]_x)+ \theta [F_1 , G_1]_x \left[2\frac{P_1}{P_0} ,H_1\right]_x\right] +\cir, 
\end{split}
\eeq
where we made use of the Jacobi identity applied to $H_1$ ,$F_1$ and $G_0$ in order to obtain the above coefficient for $ - \theta + P_0 \theta '$.

In (\ref{the}) the coefficients of the terms involving $H_1$, $F_1$ and $G_0$ must vanish separately. Such contributions correspond to the first two terms of the second integral in Eq. (\ref{the}) and to the corresponding terms obtained from cyclic permutations. We can rewrite such terms as
\beq
\begin{split}
& \int d^3 x \left[ (- \theta  + P_0 \theta ')[[H_1 , F_1 ]_x , G_0]_x +P_0 [F_1 , G_1 ]_x [\theta ' ,H_0]_x \right] + \cir=\\
& \int d^3 x \left[ - [(- \theta + P_0 \theta ' ), G_0 ]_x [H_1 , F_1 ]_x + [H_1 , F_1 ]_x [P_0 \theta ' ,G_0]_x - \theta ' [H_1 , F_1 ]_x [P_0 , G_0 ]_x \right] + \cir =0,
\end{split}
\eeq
where we made use of the Leibniz identity and of the property $ \int d^3 x [ f,g]_x =0$, for $f$ and $g$ such that boundary terms vanish when integrating by parts. 

We remark then that the coefficient of $P_1 \theta '$ in Eq. (\ref{the}) vanishes because of the Jacobi identity for $[ , ]_x$. 

The integral involving the coefficient of $P_1$, yields 
\beq
\begin{split}
& \int d^3 x P_1 \left[ [F_1 , G_1 ]_x [\theta ' , H_1]_x \right] + \cir, \\
& =\int d^3 x P_1 \left[ \pax F_1 \pay G_1 \pax \theta ' \pay H_1 - \pax F_1 \pay G_1 \pay \theta ' \pax H_1 \right. \\
& \left. - \pay F_1 \pax G_1 \pax \theta ' \pay H_1 + \pay F_1 \pax G_1 \pay \theta ' \pax H_1 \right] + \cir  =0,
\end{split}
\eeq
for any $\theta (P_0)$. Analogously, the integral involving the coefficients of $\theta$ in Eq. (\ref{the}) is identically zero. Consequently, we observe that all the terms in the Jacobiator (\ref{the}) that do not involve derivatives along $z$, vanish identically without imposing restrictions on $\theta$. The contributions including $z$-derivatives, on the other hand, enforce
\beq  \label{jacob}
\begin{split}
& \int d^3 x \left[ \theta ( G_1 [\paz F_1 ,H_1 ]_x +\paz F_1 [G_1 , H_1 ]_x -F_1 [\paz G_1 ,H_1 ]_x -\paz G_1 [F_1 , H_1 ]_x) \right. \\
& \left. + P_0( H_1 \paz \theta ' [F_1 , G_1 ]_x + H_1 \theta ' [\paz F_1 ,G_1 ]_x + H_1 \theta ' [F_1 , \paz G_1 ]_x )\right]  + \cir\\
& = \int d^3 x \left[ \theta (2 \paz H_1 [F_1 , G_1 ]_x - H_1 \paz [ F_1 , G_1 ]_x) \right.\\
 & \left. - \theta ' \paz (P_0 H_1 ) [F_1 , G_1 ]_x \right] + \cir \\
 & = \int d^3 x  (3 \theta -P_0 \theta ')  [ F_1 , G_1 ]_x \paz H_1  + \cir =0
 \end{split}
 \eeq
 for arbitrary $F$, $G$ and $H$. If one chooses $F=\int d^3 x P_1 f(x)$, $G=\int d^3 x P_1 g(y)$ and $H=\int dx' dy' dz \delta ( x -x') \delta (y - y') h(z)$, where $f$, $g$ and $h$ satisfy the required boundary conditions, then Eq. (\ref{jacob}) implies
\beq
- f'(x) g'(y) \int dz h(z) \paz (3 \theta -P_0 \theta ')  =0.
\eeq
Because this has to be valid for any $h$, we obtain that $3 \theta   -P_0 \theta '$ cannot depend on $z$. The same argument can be applied permuting the dependence of $f$, $g$ and $h$ on $x$, $y$ and $z$, which finally implies $P_0 \theta ' - 3 \theta = \Theta $ where $\Theta$ is a constant (recall that we are considering functionals with functional derivatives that do not yield finite boundary terms when integrating by parts, so that $\int d^3 x [ F_1 , G_1 ]_x \paz H_1 + \cir =0 , \quad \forall F,G, H$). Consequently, having $P_0 >0$, we obtain 
 \beq 
 \theta (P_0) = \cala P_0^3 - \frac{\Theta}{3},
 \eeq
 with $\cala$ an arbitrary constant. Thus, the Hamiltonian closure for the fluid model derived from the drift-kinetic model reads
 \beq  \label{finhamclos}
 P_2=\calf (P_0 , P_1) = \frac{P_1^2}{P_0} + \cala P_0^3 - \frac{\Theta}{3}.
 \eeq
 Because the constant $\Theta$ eventually does not give a finite contribution when inserting the expression (\ref{finhamclos}) into the bracket (\ref{bk01}), one can set $\Theta=0$ without loss of generality. 
 
It is worth remarking that, for instance, the isothermal closure $P_2=P_1^2 / P_0 + TP_0$, with constant temperature $T$ is not a Hamiltonian closure. Indeed, this closure yields
\beq
\{\{F,G\},H\} + \cir = 2T \int d^3 x  P_0  [F_1 , G_1]_x  \paz H_1 + \cir \neq 0,
\eeq
for generic $F$, $G$ and $H$.

\section{Derivation of the Hamiltonian closure for the perturbed drift-kinetic system}  \label{appb}

We look for functions $\tilde{\calf}(\tpo , \tpu)$ such that the bracket
\beq  \label{bkd}
\begin{split}
&\{F,G\}=\int d^3 x \left[ \tpo [F_0 , G_0]_x + \tpu ([F_1 , G_0]_x +[F_0 , G_1]_x) + \tfc [F_1 , G_1]_x \right.\\
& \left. -A n_0 (F_1\pa_z G_0 + F_0  \pa_z G_1 ) -2A M_0 F_1 \pa_z G_1\right],
\end{split}
\eeq
satisfies the Jacobi identity. We proceed by analogy with \ref{appa} and split the bracket (\ref{bkd}) into two contributions:
\beq
\{F,G\}=\{F,G\}_{\tc}+ \{F,G\}_{\thot},
\eeq
with
\ben
\{F,G\}_{\tc}=\int d^3 x \left[  \tpo [F_0 , G_0]_x + \tpu ([F_1 , G_0]_x +[F_0 , G_1]_x) + \right.  \nonumber \\
\left.  -A n_0 (F_1\pa_z G_0 + F_0  \pa_z G_1 ) -2A M_0 F_1 \pa_z G_1\right],  \nonumber \\
\{F,G\}_{\thot}=\int d^3 x \tfc [F_1 , G_1]_x. \nonumber 
\een
The bracket $\{F,G\}_{\tc}$ is then further decomposed as
\beq
\{F ,G \}_{\tc}=\{F ,G\}_{\tx}+\{F, G\}_{\tvl},
\eeq
where
\ben
\{F,G\}_{\tx}=\int d^3 x \left[ \tpo [F_0 , G_0]_x + \tpu ([F_1 , G_0]_x +[F_0 , G_1]_x)\right],\\
\{F,G\}_{\tvl}=\int d^3 x A \left[-n_0 (F_1 \pa_z G_0 + F_0 \pa_z G_1)-2 M_0  F_1 \pa_z G_1 \right].
\een
Given that
\ben \nonumber
(\{ F,G\}_{\tx})_0=[F_0 , G_0]_x +\text{s.v.t.}, \quad (\{ F,G\}_{\tx})_1=[F_1 , G_0]_x + [F_0 , G_1]_x +\text{s.v.t.}, \nonumber  \\  
(\{ F,G\}_{\tvl})_0=\text{s.v.t.}, \quad (\{ F,G\}_{\tvl})_1=\text{s.v.t.}, \nonumber 
\een
one obtains that
\beq  \label{bcc}
\begin{split}
&\{\{F,G\}_{\tc} ,H\}_{\tc} + \cir = \{\{F,G\}_{\tx} ,H\}_{\tvl} + \cir \\
&= 2A M_0 \int d^3 x [G_1 , H_1 ]_x \pa_z F_0 +\cir.
\end{split}
\eeq
On the other hand, one has also
\ben
(\{F,G\}_{\tc})_0=[F_0 , G_0]_x +\text{s.v.t.}, \quad (\{ F,G\}_{\tx})_1=[F_1 , G_0]_x + [F_0 , G_1]_x +\text{s.v.t.}, \nonumber \\
(\{F,G\}_{\thot})_0=\pa_0 \tfc [F_1 ,G_1]_x +\text{s.v.t.}, \quad (\{F,G\}_{\thot})_1=\pa_1 \tfc [F_1 ,G_1]_x , +\text{s.v.t.}  \nonumber
\een
and consequently
\ben  \label{bhh}
\{ \{F,G\}_{\thot} , H \}_{\thot}  + \cir= \int d^3 x \tfc [F_1 , G_1]_x [ \pa_1 \tfc ,H_1]_x +\cir , 
\een
\ben  \label{bch}
\{ \{F,G\}_{\tc} , H \}_{\thot} + \cir =\int d^3 x \tfc [[F_0 , G_1]_x + [F_1 ,G_0]_x , H_1 ]_x + \cir, 
\een
\beq    \label{bhc}
\begin{split} 
& \{ \{F,G\}_{\thot} , H \}_{\tc} + \cir=\int d^3 x \left[ \tpo [\pa_0 \tfc [F_1 , G_1]_x ,H_0]_x \right.\\
&\left. +\tpu ( [ \pa_0 \tfc [F_1 , G_1 ]_x , H_1 ]_x + [ \pa_1 \tfc [F_1 , G_1 ]_x ,H_0]_x ) \right. \\
&\left.  - A n_0 (\pa_1 \tfc [F_1 , G_1 ]_x \pa_z H_0 + \pa_0 \tfc [F_1 , G_1 ]_x \pa_z H_1 )  - 2 A M_0 \pa_1 \tfc [F_1 , G_1]_x \pa_z H_1 \right] + \cir.
\end{split}
\eeq
Combining (\ref{bcc}), (\ref{bhh}), (\ref{bch}) and (\ref{bhc}) one obtains 
\beq  \label{jacdf}
\begin{split}
&\{ \{F,G \} ,H\} + \cir  \\
&= \{\{F,G\}_{\tc} ,H\}_{\tc} + \{ \{F,G\}_{\thot} , H \}_{\thot} + \{ \{F,G\}_{\tc} , H \}_{\thot} +  \{ \{F,G\}_{\thot} , H \}_{\tc} + \cir\\
&=\int d^3 x \left[2A M_0 [G_1 , H_1 ]_x \pa_z F_0 + \tpo [\pa_0 \tfc [F_1 , G_1]_x ,H_0]_x  \right.\\ 
&\left. + \tpu ([ \pa_0 \tfc [F_1 , G_1 ]_x , H_1 ]_x +[\pa_1 \tfc [F_1 , G_1 ]_x , H_0]_x ) \right.\\
&\left. -A n_0 (\pa_1 \tfc [F_1 ,G_1 ]_x \pa_z H_0 + \pa_0 \tfc [F_1 , G_1 ]_x \pa_z H_1 ) \right. \\
&\left. -2A M_0 \pa_1 \tfc [F_1 ,G_1 ]_x \pa_z H_1 + \tfc [F_1 , G_1]_x [\pa_1 \tfc ,H_1 ]_x \right. \\
&\left. +\tfc ([[F_0 ,G_1]_x ,H_1 ]_x +[[F_1 ,G_0]_x , H_1 ]_x )\right] +\cir .
\end{split}
\eeq
Similarly to what occurred in the case of the full drift-kinetic system, also in the perturbed system, imposing the Jacobi identity, implies a cancellation between the only two terms in (\ref{jacdf}) involving $G_1$, $H_1$ and $\pa_z F_0$ (and of course, similarly for the other terms obtained from those by means of cyclic permutations). This leads to the constraint $n_0 \pa_1 \tfc =2 M_0$, which implies
\beq  \label{forfdf}
\tfc (\tpo , \tpu)= 2\frac{M_0}{n_0} \tpu +\gamma(\tpo),
\eeq
with $\gamma$ an arbitrary function. Inserting the form  (\ref{forfdf}) into (\ref{jacdf}) and using the Leibniz identity and the Jacobi identity for $[ , ]_x$, yields
\beq \label{jacdf2}
\begin{split}
&\{ \{F,G \} ,H\} + \cir  \\
&= \int d^3 x \left[ \tpo [ \gamma ' [F_1 , G_1 ]_x ,H_0]_x -A n_0 \gamma ' [F_1 , G_1 ]_x \pa_z H_1 \right.\\
& \left. +\gamma  ([[F_0 ,G_1]_x ,H_1 ]_x +[[F_1 ,G_0]_x , H_1 ]_x )  + \tpu [ \gamma ' [F_1 , G_1 ]_x , H_1 ]_x \right] +\cir .
\end{split}
\eeq
In (\ref{jacdf2}), the sum of the terms involving only functional derivatives with respect to $\tpu$ have to vanish independently. Again, making use of the Leibniz identity, this implies
\beq  \label{iddf}
A n_0 \int d^3 x F_1 (\pa_z G_1 [H_1 , \gamma ']_x - \pa_z H_1 [G_1 , \gamma ']_x )=0.
\eeq
Because (\ref{iddf}) has to be valid for arbitrary $F$, $G$ and $H$, this leads to $\gamma=a P_0 + \Gamma$, with constant $a$ and $\Gamma$. It turns out that with this solution for $\gamma$, also all the remaining terms in Eq. (\ref{jacdf2}) vanish, by virtue of the Jacobi identity for the bracket $[ , ]_x$. Recalling Eq. (\ref{forfdf}), we obtain then that the final form for $\tfc$ is given by
\beq  \label{clolin}
\tfc (\tpo , \tpu)= a \tpo +   2\frac{M_0}{n_0} \tpu,
\eeq
where, analogously to \ref{appa}, we set equal to zero the irrelevant constant $\Gamma$.

\end{document}